\documentclass[9pt,twocolumn,twoside]{pnas-new}

\templatetype{pnasresearcharticle}

\title{A data-driven framework for structure-property correlation in ordered and disordered cellular metamaterials}

\author[a]{Shengzhi Luan}
\author[a]{Enze Chen} 
\author[a]{Joel John}
\author[a,1]{Stavros Gaitanaros}

\affil[a]{Department of Civil and Systems Engineering, Johns Hopkins University, Baltimore, MD 21218}

\leadauthor{Luan} 

\correspondingauthor{\textsuperscript{1}To whom correspondence should be addressed. E-mail: stavrosg@jhu.edu}


\begin{abstract}
Cellular solids and micro-lattices are a class of lightweight architected materials that have been established for their unique mechanical, thermal, and acoustic properties. It has been shown that by tuning material architecture, a combination of topology and solid(s) distribution, one can design new material systems, also known as metamaterials, with superior performance compared to conventional monolithic solids. Despite the continuously growing complexity of synthesized microstructures, mainly enabled by developments in additive manufacturing, correlating their morphological characteristics to the resulting material properties has not advanced equally. This work aims to develop a systematic data-driven framework that is capable of identifying all key microstructural characteristics and evaluating their effect on a target material property. The framework relies on integrating virtual structure generation and quantification algorithms with interpretable surrogate models. The effectiveness of the proposed approach is demonstrated by analyzing the effective stiffness of a broad class of two-dimensional (2D) cellular metamaterials with varying topological disorder. The results reveal the complex manner in which well-known stiffness contributors, including nodal connectivity, cooperate with often-overlooked microstructural features such as strut orientation, to determine macroscopic material behavior. We further re-examine Maxwell's criteria regarding the rigidity of frame structures, as they pertain to the effective stiffness of cellular solids and showcase microstructures that violate them. This framework can be used for structure-property correlation in different classes of metamaterials as well as the discovery of novel architectures with tailored combinations of material properties.
\end{abstract}

\begin{document}

\maketitle
\thispagestyle{firststyle}
\ifthenelse{\boolean{shortarticle}}{\ifthenelse{\boolean{singlecolumn}}{\abscontentformatted}{\abscontent}}{}

% If your first paragraph (i.e. with the \dropcap) contains a list environment (quote, quotation, theorem, definition, enumerate, itemize...), the line after the list may have some extra indentation. If this is the case, add \parshape=0 to the end of the list environment.

\section*{Introduction}

Ongoing advances in additive manufacturing have led to a rapid growth on the synthesis of architected materials with increasingly complex microstructures across several length scales \cite{Fleck_2010_PRSA,Schaedler_2016_ARMR}. Cellular metamaterials, consisting of polyhedral topologies with struts \cite{Fleck_2006_JMPS,Zheng_2016_NatMat}, plates \cite{Lorenzo_2020_NatCom,Fang_2020_ActMat}, or shells \cite{Han_2015_AdvMat,Lorenzo_2021_ComStr,Gao_2022_PNAS} as building blocks, are an important class of architected materials that has been the focus of a vast number of studies due to their unique mechanical, thermal and acoustic properties and their ubiquitous occurrence in nature \cite{Gibson_Book_2010}. To date, a plethora of engineered architectures have been reported with desirable effective properties including high stiffness \cite{Zheng_2014_Science,Gurtner_2014_PRSA,McMeeking_2017_Nature,Dirk_2018_AdvMat}, negative Poisson's ratio \cite{Lakes_1987_Science,Katia_2010_AdvMat,Lewis_2015_AdvMat}, phononic bandgaps \cite{Fleck_2006_JASA,Stavros_2018_JAM}, energy absorption \cite{Carlos_2021_NatMat,Guest_2022_ScrMat,Dirk_2016_ActMat}, heat transfer \cite{Catchpole_2019_AM}, strength and resilience \cite{Greer_2011_Science,Bauer_2014_PNAS,Bauer_2016_NatMat,Meze_2015_PNAS,Chiara_2019_PNAS}, toughness \cite{Deshpande_2022_NatMat,Luan_2022_JMPS}, and others. Rigorous design methodologies, such as topology optimization, have led to material systems with stiffness and strength that reach their individual theoretical bounds \cite{Gurtner_2014_PRSA}. Recently, data-driven techniques have also been employed \cite{Dennis_2022_PNAS,Lumpe_2021_PNAS,Xuanhe_2020_SciAdv} in order to further explore the property space of architected materials.

Despite the significant developments in both design algorithms and synthesis techniques, the analysis of the resulting material mechanics, and in particular the correlation of effective properties to all critical morphological features, has not progressed equally. Experimental measurements of truss-based metamaterials are often interpreted through the lens of classical scaling formulas, derived from dimensional analysis and beam theory. In this approach, an effective material property $\Bar{y}$ is connected with the underlying microstructure through a power law i.e., $\Bar{y} = \alpha\Bar{\rho}^\beta$, where $\Bar{\rho}$ is the relative density of the architected material, and $(\alpha,\beta)$ are constants that depend on the properties of the parent solid and all topological characteristics. Typically, the exponent $\beta$ is determined by the axial-to-bending ratio of the internal loads within the struts, designating the metamaterial as either stretching- or bending-dominated. This classification, in turn, is related to the corresponding average nodal connectivity. Several works \cite{Carlos_2018_EML,Meza_2017_ActMat,Enze_2022_IJSS} have illustrated the limitations of these scaling laws (e.g. not accounting for shear and/or the complex effect of the strut junctions) and reported mixed results regarding their agreement with experimental measurements on different microstructures. However, to date, a framework which enables comprehensive and accurate structure-property correlation, and at the same time is sufficiently general to be applicable to broad classes of architected materials, remains an open challenge.

In this work, we demonstrate a systematic data-driven approach that exploits cellular structure generation and quantification algorithms, to yield interpretable regression-based surrogate models with the ability to not only accurately capture a target material property, but more importantly identify key morphological descriptors and evaluate their effect on macroscopic material behavior. The accuracy and effectiveness of the proposed framework are demonstrated by analyzing the effective stiffness of 2D cellular metamaterials and elucidating the complex manner on which individual microstructural descriptors contribute to it. A distinct element of the techniques developed here, is that they enable the simultaneous study of both perfectly-ordered lattices and highly disordered microstructures. It is also important to emphasize that all aspects of individual components of this framework (i.e. virtual structure generation, microstructure quantification, property estimation, and structure-property correlation) can be extended to other classes of metamaterials and various properties of interest.

\section*{Virtual Structure Generation}

In order to generate a large, but most importantly representative, dataset of truss metamaterials we introduce a virtual structure generation framework that combines tiling patterns and power diagrams. Our objective is to create cellular architectures with a varying, and to an extent tailored, amount of disorder so that the resulting materials, in turn, attain a wide range of properties. Power diagrams, also known as Laguerre diagrams or Dirichlet cell complexes, have found applications in many scientific disciplines including solid state physics, geology, and economics \cite{Rimoli_2020_CMAME,Aurenhammer_1987_SIAM}. A finite-size planar domain $M$ is first divided into a set $\boldsymbol{S}\in\mathbb{R}^2$ of $n$ cells, each associated with a seed $\boldsymbol{s}_i\in\boldsymbol{S}=\{\boldsymbol{s}_1,\boldsymbol{s}_2,\cdot\cdot\cdot\boldsymbol{s}_n\}$ that is located at $\boldsymbol{c}_i$. The convex region $C$ occupied by a single cell is defined by 
\begin{equation}
    C_i = \Bigl\{\boldsymbol{x}\in\mathbb{R}^2 \mid d_L(\boldsymbol{x},\boldsymbol{s}_i)<d_L(\boldsymbol{x},\boldsymbol{s}_j), \forall\boldsymbol{s}_j\in\boldsymbol{S}-\{\boldsymbol{s}_i\}\Bigl\}.
\end{equation}
where $d_L(\boldsymbol{x},\boldsymbol{s}_i)=\lVert \boldsymbol{x}-\boldsymbol{c}_i \rVert^2-w_i^2$ is the Laguerre (or power) distance and $w_i$ is the weight assigned to each seed. A common geometric interpretation of a seed with weight is a circle with radius $w_i\geq0$. By controlling the placement of seeds and their corresponding weights, one can design innumerable polygonal tessellations. Note that in classical Voronoi diagrams all weights are equal and $d_L$ is replaced by the Euclidean distance. Both Voronoi and Laguerre tessellations have been used extensively as representations of cellular microstructures found in nature, such as plant tissue and trabecular bone \cite{Gibson_Book_2010}.

The structure generation framework proposed here follows a two-step process. In the first step, we design a primitive set of cellular metamaterials that encompasses, though not exclusively, well-known 2D architectures that have been the foci of numerous research works due to their unique material properties. Since a large portion of these materials consists of periodic microstructures, we employ tilings of the 2D Euclidean space as blueprints, in order to generate seed locations $\boldsymbol{c}_i$ that yield perfectly-ordered topologies with distinct geometric characteristics. The art of tilings i.e., filling space with unit-tiles of a specific shape, can be traced back to antiquity \cite{Grunbaum_Book_1987}, and more recently to the work of M.C.Escher \cite{Escher_Book_2000}. Although several of their geometric properties were known to ancient Greek mathematicians, a systematic analysis and listings of tilings first appear in the works of Kepler \cite{Kepler_Book_1969}. Recent studies have focused on discovering new families of polyhedral tilings with applications in condensed matter physics and crystal structures \cite{Conway_2011_PNAS,Friedrichs_1999_Nature}, though to date, a complete classification of 2D tilings by polygons remains an open-challenge \cite{Chavey_1989_Sym2}. In this work we consider solely the case of edge-to-edge tilings, that is if two polygons intersect at more than one point then they must share a common edge. If each polygonal cell is vertex-transitive i.e., all vertices are equivalent with respect to the corresponding symmetry group, then the tiling is \emph{uniform} (or 1-uniform). It was shown by Kepler that there are only 11 uniform tilings in $\mathbb{R}^2$. These include three regular tilings and eight semi-regular tilings that are also known as Archimedean. The former group consists of only one type of regular polygons (triangles, squares, and hexagons respectively) while the semi-regular tilings are formed by two or three types of regular polygons. By increasing the tile-transitivity one can define general classes of \emph{k-}uniform tilings \cite{Grunbaum_Book_1987}. Even though \emph{k-}uniform tilings have been enumerated up to $k=6$, here we will focus on materials based on the 1- and 2-uniform tilings, since the remaining classes do not contribute appreciably towards microstructural diversity. Of course, it is impossible to represent a broad class of ordered lattice metamaterials using only combinations of regular polygons, irrespective of the range of their types. To address this issue, we take advantage of the dual tilings, which are generated by placing seeds in the vertices of the polygons, and then tessellate the same planar domain with single-valued weights (Supporting Information, Fig. S1). Subsequently, we add to this set of designs two disordered topologies: a Voronoi diagram generated by seed placement based on a uniform distribution, and its corresponding dual tessellation.

\begin{figure*}[t]
\centering
\includegraphics[width=1.0\linewidth]{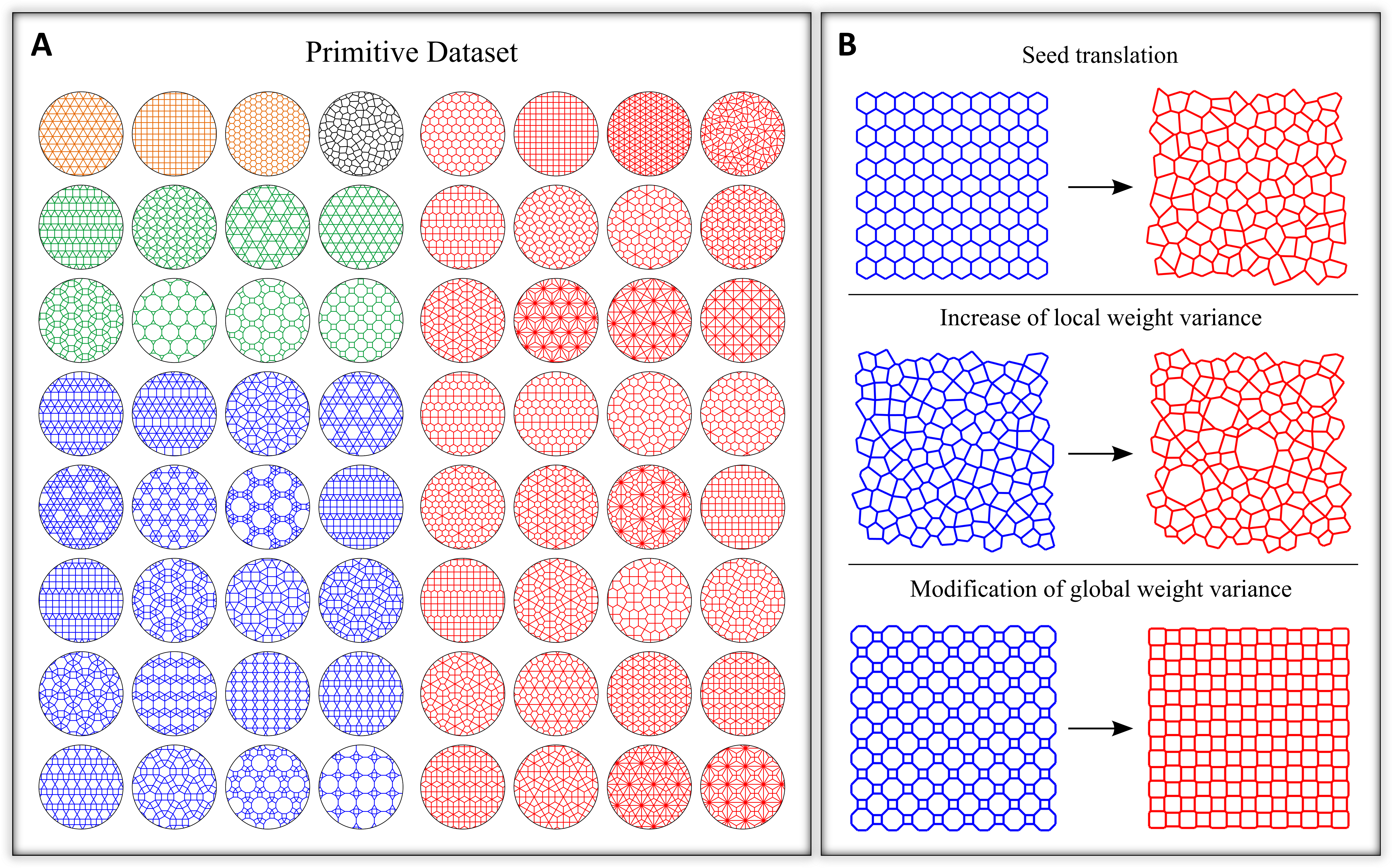}
\caption{Basic cellular architectures and structure modification. (A) Primitive set of 64 architectures that includes: eleven 1-uniform tilings (three regular marked in yellow and eight semi-regular marked in green), twenty 2-uniform tilings (marked in blue), a disordered Voronoi structure (marked in black), and their corresponding dual tessellations (marked in red). (B) Three operations (seed translation, increase of local weight variance, and modification of global weight variance) are used to expand the primitive set to an expanded representative dataset of 1646 architectures with wide distributions of morphological characteristics.}
\label{Figure:F1}
\end{figure*}

The process described above leads to 64 generated architectures (see Fig. \ref{Figure:F1}A), of which 60 are distinct (4 of the dual tilings correspond to an existing topology), that will form the basis for an expanded set of microstructures with much richer morphologies. It is important to emphasize here that this primitive dataset of architectures includes most, if not all, well-studied 2D cellular solids \cite{Fleck_2010_PRSA,Gurtner_2014_PRSA} including the triangular, square and hexagonal honeycombs, the Kagome lattice, a subset of the stiffest isotropic lattices, and an archetypal random material. The second step in the microstructure generation process involves the expansion of this primitive set, by modifying the seed locations and their weights, to a large database of 2D structures with an amplifying amount of disorder. Three distinct operations (Fig. \ref{Figure:F1}B) are employed to generate the final expanded set of 2D structures from each one of the basic designs: (a) seed translation, (b) increase of local weight variance, and (c) modification of global weight variance. The goal of the first operation is twofold. First, it allows us to break the symmetry of the basic periodic topologies, leading to a large group of microstructures with distinct disorder characteristics than the typical Voronoi tessellations. Second, it enables assessing the effect of defects or imperfections, on the resulting macroscopic properties, which is key to the design of robust architected metamaterials. The two operations that affect the weight of the seeds lead to individual distributions of cell sizes and thus, to an additional group of disordered metamaterials that also differs from random seed placement-based disordered topologies. It is important to emphasize that not all three operations are applicable, or meaningful, for all microstructures contained in the primitive set.

Seed translation refers to the change of seed position from its corresponding value in the initial primitive design (Supporting Information, Fig. S2a). For every seed $\boldsymbol{s}_i$, its position $\boldsymbol{c}_i$ is modified by a radial distance $r$ along a random direction $\theta$, resulting in a updated seed position $\boldsymbol{c}'_i$ with
\begin{equation}
    \boldsymbol{c}'_i = \boldsymbol{c}_i+r[\cos{\theta},\sin{\theta}]^\text{T}.
\end{equation}
The two uniform random variables $r$ and $\theta$ are generated independently within the ranges $r\sim U(0,\xi_\text{p}d_\text{s})$ and $\theta\sim U(0,2\pi)$ respectively. The coefficient $\xi_\text{p}$ corresponds to the maximum relative change of position, whereas $d_\text{s}$ denotes the minimum distance between neighboring seeds. Seed translation is only applied to the regular microstructures of the basic set i.e., the 1- and 2-uniform tilings and their duals, since this operation has a minimal effect on the microstructural characteristics of the disordered Voronoi tessellations.

Local weight variance describes the fluctuation of the seed weight from its original value in the basic tessellations. Here, each seed with weight $w_i$ is replaced by a new seed with weight $w'_i=\eta_\text{v}w_i$, where $\eta_\text{v}\sim N(1,\xi_\text{v})$ follows a normal distribution (Supporting Information, Fig. S2b) scaled by the local weight coefficient $\xi_v$. Increasing the local weight variance is applied to all designs included in the primitive set. Global weight variance refers to the macroscopic distribution of seed weights with respect to their mean value $\bar{w}$ (Supporting Information, Fig. S2c). Using a global weight coefficient $\xi_\text{d}$, each seed $\boldsymbol{s}_i$ is assigned a new weight $w'_i$ according to 
\begin{equation}
    w'_i = w_i+\xi_\text{d}(\bar{w}-w_i).
\end{equation}
Modifying global weight variance is only meaningful on designs generated by power diagrams with non-uniform weights, that is, the semi-regular and 2-uniform tilings. It is trivial to show that this operation leaves all other designs of our primitive set intact.

Applying these structure modification techniques to the basic set of 2D materials leads to a broader class of cellular topologies consisting of 1646 designs. Finally, to transition from pure geometric entities to real metamaterials, one needs to assign a solid material distribution along the polygonal topologies. In general, there is an infinite number of ways to achieve this by exploiting different combinations of non-uniform density distributions with arbitrary shapes of cross-sectional areas for each strut. For simplicity and tractability, here we assign uniform strut thickness to all edges of the underlying tessellations in order to attain a target relative density of the resulting truss metamaterial. To avoid errors in the relative density calculation, material overlap at the strut-junctions is removed. For each tessellation, we assign five different values of strut thickness to obtain corresponding relative densities $\bar{\rho} = [1\%, 5\%, 10\%, 15\%, 20\%]$. Therefore, our final set of ordered and disordered cellular metamaterials consists of 8230 distinct microstructures.

\begin{figure*}[b]
\centering
\includegraphics[width=1.0\linewidth]{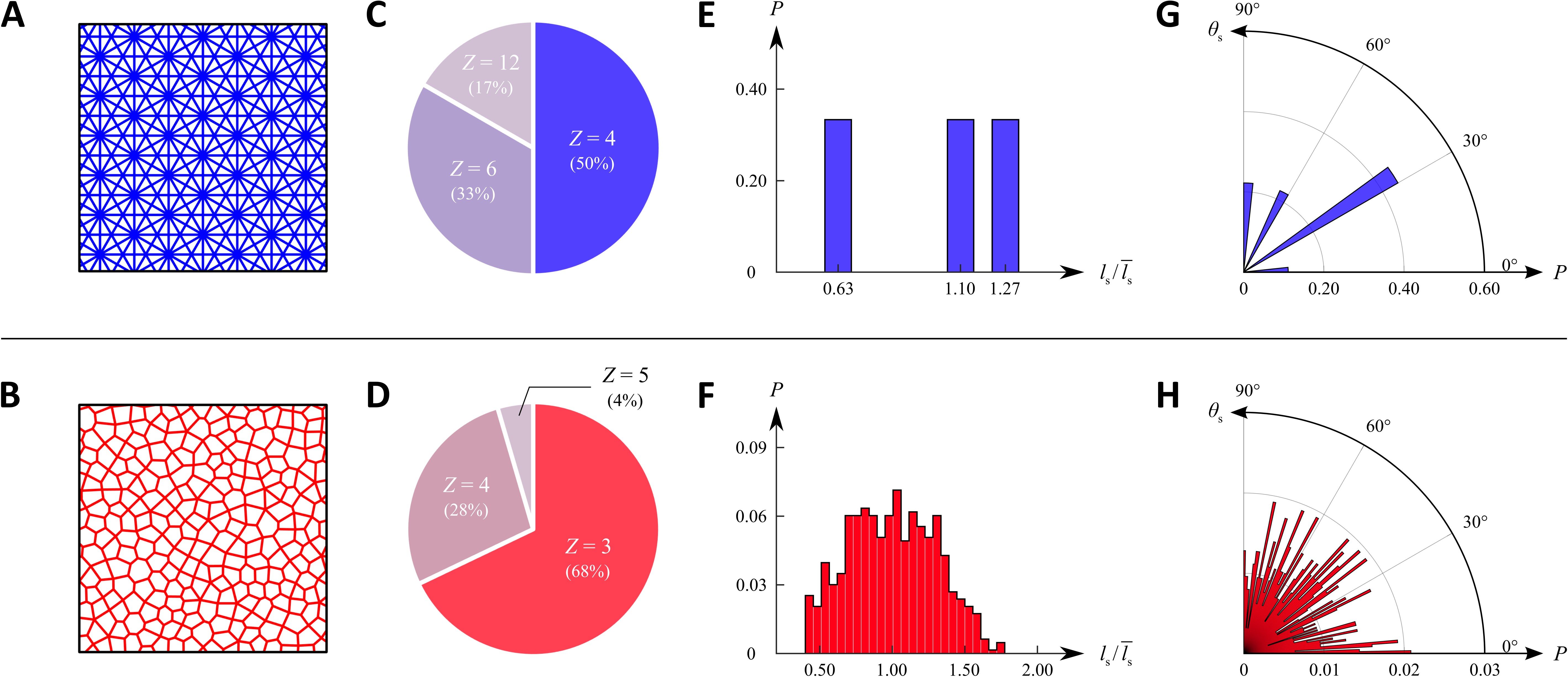}
\caption{Microscopic level descriptors for two samples: a periodic (A) and a disordered (B) metamaterial with their corresponding distributions of nodal connectivity (C,D), strut length (E,F), and strut orientation (G,H).}
\label{Figure:F2}
\end{figure*}

\section*{Microstructure Quantification}

An essential component in the analysis of cellular metamaterials is an accurate description of their microstructural properties. To date, this is typically based on geometric classifications (e.g. periodic, three-dimensional, etc.), underlying crystal symmetries (e.g. FCC), or the mechanical behavior at the strut level (e.g. stretching-dominated). However, none of the above approaches is complete, failing to properly capture the complexity of cellular architectures, especially non-periodic ones. Recently, image-based tools have been employed \cite{Xuanhe_2020_SciAdv} to address this issue and re-explore the design space of lightweight architected materials. Here we aim to establish a rigorous framework, employing a set of well-established metrics of morphological features, that allows the digitalization of microstructure characterization. The main advantage of this approach is that it can facilitate comparisons between different metamaterials and more importantly enable a systematic construction of structure-property relations, as is shown in the next section.

Based on previous studies focusing on identifying key microstructural features and their effects on the mechanical properties of porous, polycrystalline, and multi-phase composites, we adopt here a vector representation of cellular metamaterials, where each component represents a microstructural feature \cite{WKLiu_2018_PMS}. At the metamaterial (macroscopic) scale, microstructure is characterized by the relative density (or volume fraction) and number of polygonal cells. Each cell is in turn described by its edge number, area, compactness and eccentricity. At a microscopic level, features include each strut length, orientation and nodal connectivity (also known as coordination number in solid state physics). Connecting these length-scales are the distance and angle between the seeds of neighboring cells. Note that we calculate the sine/cosine of all angles for the seeds to avoid any periodicity-induced bias issues. All of the above characteristics (Supporting Information, section 2) can be classified as either deterministic i.e., denoting a property (e.g. relative density) of the entire structure by a single value, or statistical, representing a given feature (e.g. strut length) by its probability distribution function across the material domain. For each feature in the latter category, we employ four descriptors corresponding to the first four moments of their probability distribution function i.e., its mean, variance, skewness and kurtosis. This process leads to a material descriptor vector $\mathbf{C}$ with 42 components. The microscopic level features are shown in Fig. \ref{Figure:F2} as an example, while a detailed report for all structure components is included in Supporting Information, Fig. S3. Collectively, these morphological descriptors are tailored to the specific, even though broad, class of polygonal cellular microstructures examined in this study. However, different or additional characteristics can be used for different types of architected materials.

At this point it is important, with the aid of our microstructure quantification framework, to re-examine how representative the set of generated metamaterials is, and further discuss the comprehensiveness of the microstructure vector $\mathbf{C}$. A potential pitfall in using select physical descriptors to characterize a given microstructure is the possibility of over-simplifying the representation of the material and creating microstructural uncertainty \cite{Miguel_2017_CMAME}. This in turn, can result in unfeasible or erroneous structure-property correlation. Using the four moments of all statistical morphological features, however, promotes specificity of the corresponding probability distributions, since higher moments typically do not offer additional information. The large number of adopted descriptors and their intercorrelation indicate that it is improbable, if not impossible, to generate two distinct materials with the exact same sets of descriptors. Furthermore, each sub-set of materials (e.g. ordered) included in our general class of metamaterials should be uniquely represented by these descriptors. This can be seen, for example, in Fig. \ref{Figure:F2}, where it is obvious that a periodic lattice will result in a discrete and discontinuous distribution of strut orientations, while the corresponding distribution of a disordered metamaterial is continuous within the same range. Note that these differences are consistent across many different descriptors. Finally, we ensure that the range of values for each microstructural feature, once recorded from all materials, is sufficiently extensive. This serves as additional evidence that our material dataset constitutes a representative set of 2D cellular microstructures (Supporting Information, Fig. S4). It is important to note that this result is not coincidental but was purposefully ensured by modifying all parameters of the structure generation process described in the previous section in an iterative process. In particular, the magnitudes of the three topological variation operations induced in the primitive set of material topologies were tuned to achieve the desired amount of microstructural richness.

\section*{Effective Stiffness Prediction}

\begin{figure}[h]
\centering
\includegraphics[width=1.0\linewidth]{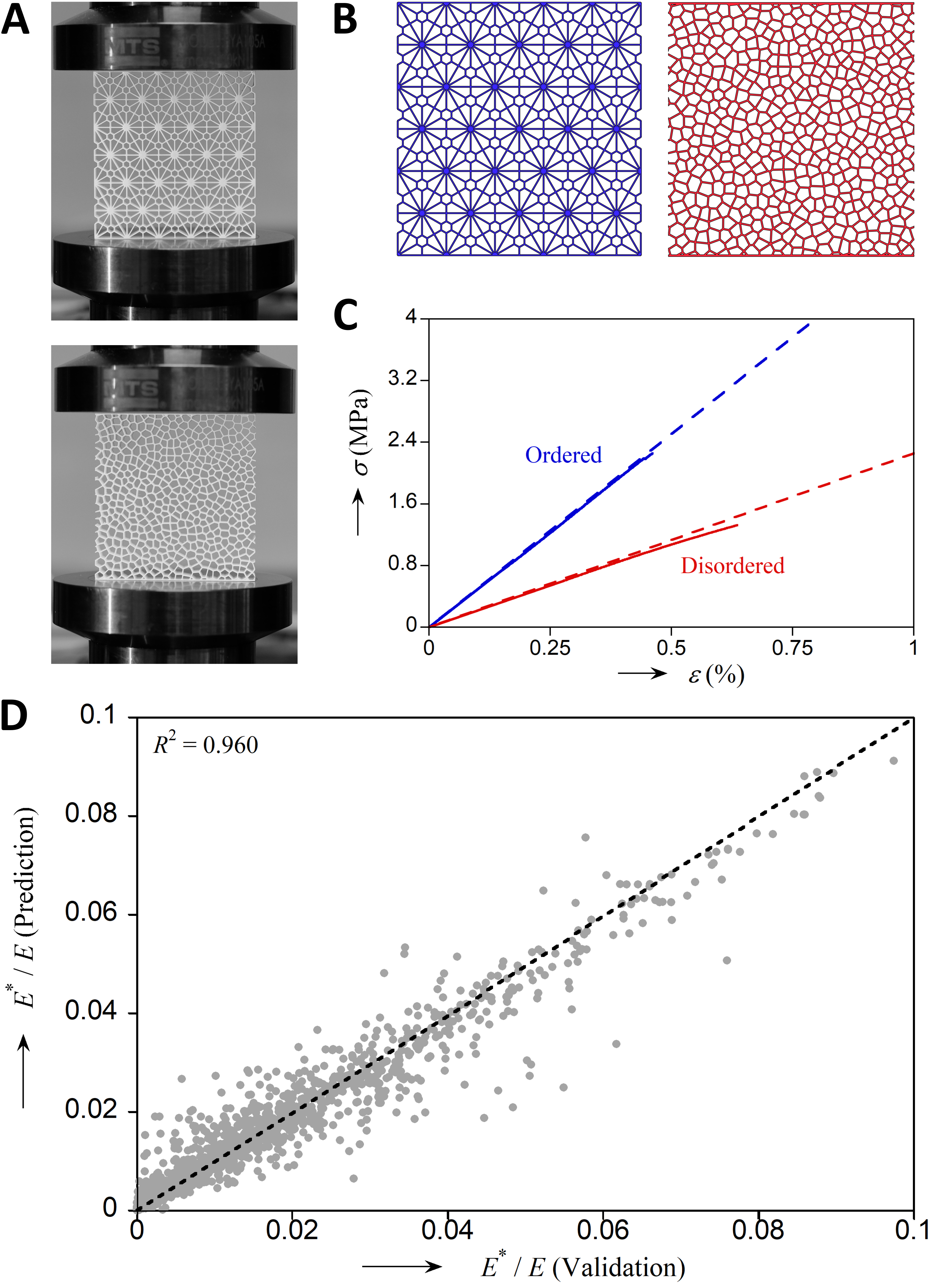}
\caption{(A) Synthesized specimens for two cellular metamaterials and (B) their corresponding numerical models. (C) Comparison of experimental (solid) and numerical (dashed) stress-strain responses for the same samples. (D) Prediction vs. validation of relative effective stiffness for the random forest model.}
\label{Figure:F3}
\end{figure}

We proceed to show that the structure vector $\mathbf{C}$ can be incorporated in machine learning algorithms to create a surrogate model that can accurately predict a material property of any 2D cellular structure. Here we focus on effective stiffness, calculated through numerical simulations due to their low computational cost and high accuracy. All material microstructures are discretized with finite elements and are subsequently compressed in silico along the vertical direction. Both ordered and disordered topologies are treated as finite-size domains i.e., there is no periodicity applied to any simulation and the transverse edges remain constraint-free. To validate this modeling framework we select two architected materials, one ordered and one disordered, with distinct microstructural vectors, synthesize them by stereolithography and test them experimentally. Fig. \ref{Figure:F3} shows the comparison between measured and predicted compressive responses for the two microstructures, demonstrating the high accuracy and efficiency of the numerical models (all details of the simulations, synthesis, and testing processes are detailed in Materials and Methods). Hence, we employ the same procedure for all metamaterials in our dataset and record their relative stiffness $\Bar{E}=E^*/E$, where $E^*$ is the effective stiffness of the metamaterial and $E$ is the Young's modulus of the parent solid.

Compiling the microstructural features and effective stiffness of each metamaterial enables the construction of a machine learning-based surrogate model with the ability to predict this target property directly from the structure vector $\mathbf{C}$. A feature selection step is applied first, using filtering techniques based on the Pearson correlation coefficient $p$ and mutual information $I$, to increase robustness, accuracy and interpretability of the surrogate model. Pearson correlation measures the linear dependency between components of the feature vector and is employed here to reduce the inherent multicollinearity of the microstructural descriptors. Two pairs of these descriptors are found to be highly correlated: the mean of nodal connectivity has a negative correlation with the mean of cell edge number ($p=-0.927$); the mean of nearest neighboring seed distance has a positive correlation with the mean of cell area ($p=0.920$). Proceeding, we retain only the average nodal connectivity and nearest neighboring seed distance in the structure vector $\mathbf{C}$. Mutual information indicates the dependency between target property and each microstructural descriptor and can therefore be used to eliminate redundant components. Here, only descriptors with $I=0$ i.e., indicating no correlation with the target variable, are removed. Through the two filtering processes, the remaining structure vector has 23 components. It is important to note that the updated, and reduced, vector $\mathbf{C}$ only applies to the material property under focus, while the original one can be applied towards any other material property.

\begin{figure*}[t]
\centering
\includegraphics[width=1.0\linewidth]{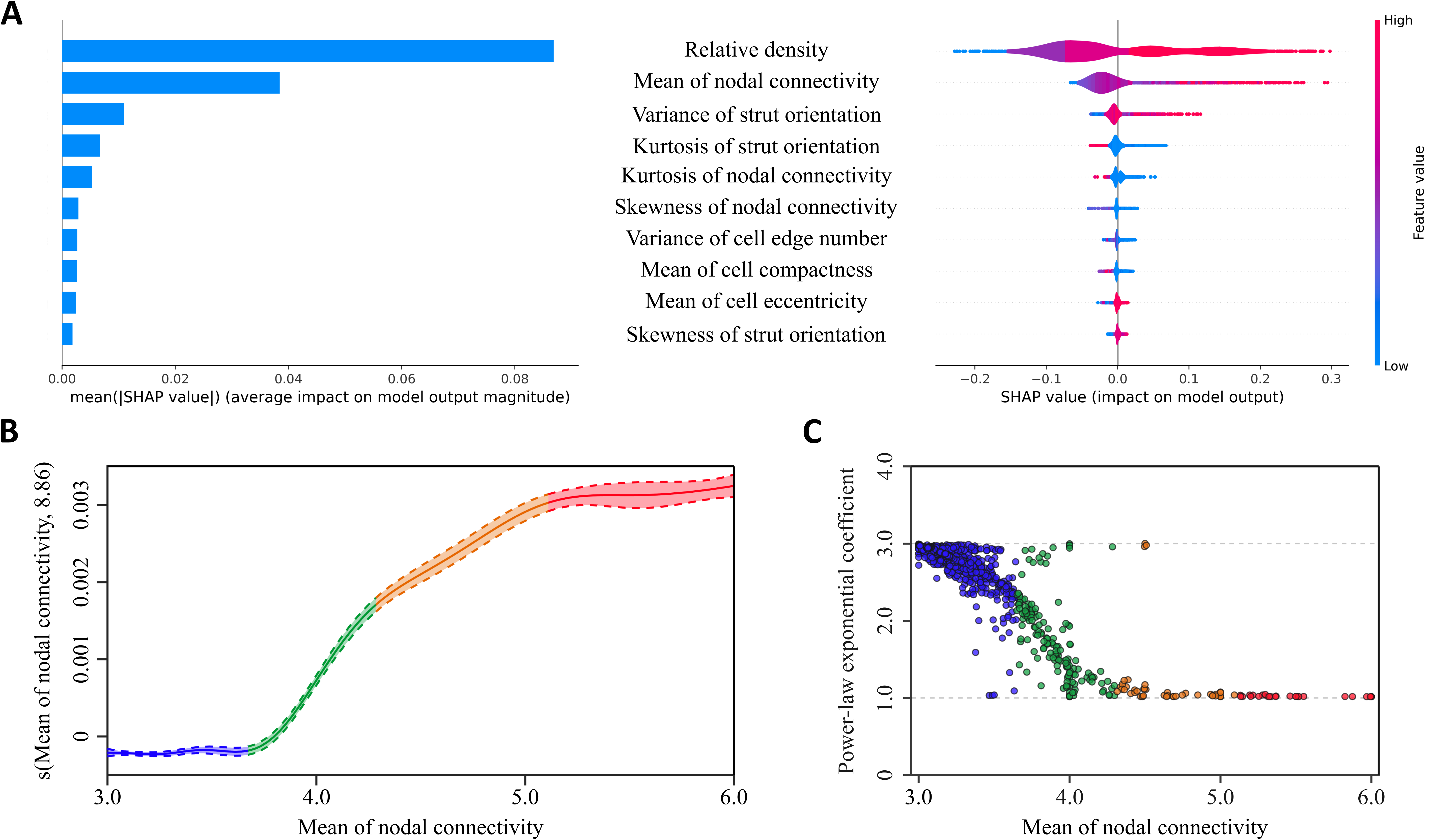}
\caption{(A) The ten most essential descriptors ranked by their overall impact, as denoted by their mean absolute Shapley value (left). The summary plot (right) combines descriptor importance with their overall effect on stiffness: the color represents the relative value of each descriptor from low (blue) to high (red), while the x-axis (SHAP value) denotes the impact on the stiffness. (B) GAM partial dependence plot showing the relationship between effective stiffness and mean nodal connectivity. (C) Power exponents of the scaling of stiffness with density for all samples in the dataset as a function of their mean nodal connectivity.}
\label{Figure:F4}
\end{figure*}

Subsequently, we construct a random forest regression model in which a number of decision trees are trained on different subsets of the material dataset, including corresponding data of effective stiffness and microstructural descriptors. Optimal parameters are determined by a grid-search method based on the mean-squared error criterion. The prediction of the resulting regression model is generated by aggregating all predictions by each individual tree. The material dataset is randomly split into 70\% for training and the remaining 30\% for testing. The predicted values of the relative effective stiffness by the surrogate model are compared with the corresponding validation data in Fig. \ref{Figure:F3}D. The effectiveness and accuracy of the machine learning algorithm are reflected in the overall satisfactory agreement and the resulting variance explained (coefficient of determination) $R^2=0.960$. Additionally, by checking predictions and validations for different samples we verify that the accuracy of the algorithm is independent of the specific class of microstructures (e.g. ordered, monodisperse, low-density etc.). It is important to note that despite the excellent performance of the surrogate model, there is an inherent limitation induced by the representation of the microstructure using physical descriptors (vs. an image-based technique for example) that leads to an inevitable loss, though limited, of structure information. However, the particular microstructure quantification and associated learning algorithm adopted here are chosen to favor interpretability, which will be critical in extracting accurate structure-property relations, as is shown next.

\section*{Structure-Property Correlation}

To uncover the intricate relation between effective stiffness and key microstructural descriptors, we employ a Shapley Additive exPlanations (SHAP) framework \cite{SHAP} customized for decision-tree models. This technique takes advantage of Shapley values, a game-theory based metric of the contribution by each player, that has recently been extensively utilized for the interpretation of machine-learning model predictions. In this context, the players are the microstructural descriptors and the game's outcome corresponds to the surrogate model's stiffness prediction. Running SHAP for all microstructures in our dataset results in an $8120\times23$ matrix of Shapley values. Fig. \ref{Figure:F4}A presents a summary plot that demonstrates the effect of each descriptor, ordered according to their importance (given space constraints, only the ten most essential descriptors are shown). As expected, the effective stiffness is governed by the relative density in a positive relation, as is the case for any porous material. The average nodal connectivity, another well-established parameter, is shown to be the second most critical factor, exhibiting an overall positive impact on the effective stiffness. Strut orientation plays a vital role as well in determining stiffness but in a non-straightforward manner. According to SHAP analysis, the third and fourth most important descriptors are the variance and kurtosis of the strut orientation distribution, in a positive and negative manner respectively. Both of these results imply that the existence of a large number of struts with deviating orientations from their mean value is highly desirable to attain high stiffness, though not necessary for all microstructures as shown next.

Furthermore, one can also use the SHAP framework to interpret individual model predictions. Doing so in select cellular architectures (for details see Supporting Information, section 3), shows that: (i) isotropic metamaterials attain the highest effective stiffness mainly due to their corresponding high nodal connectivity $Z=6$; (ii) the variance of strut orientation becomes extremely important when the average nodal connectivity is not close to its extremal values; (iii) despite the relatively low nodal connectivity,  features such as low kurtosis of strut orientation distribution help the well-studied Kagome lattice to reach a high stiffness, comparable to the corresponding rigidity of the triangular lattice.

To further probe the relation between cellular microstructure and effective stiffness, we take advantage of an additional interpretable regression technique, that is, Generalized Additive Model (GAM). GAMs enable modeling of non-linear relationships, with no strong assumption of their form, between a target response variable (i.e. effective stiffness) and one or more of the covariates (i.e. the microstructural descriptors). Here we focus solely on topological features and therefore eliminate relative density as a descriptor while keeping its value constant ($\Bar{\rho}=1\%$) for all samples. The resulting predictions have a slightly lower accuracy ($R^2=0.877$) than the corresponding ones of the random forest model, but are still sufficient to elucidate how key descriptors, as identified by SHAP analysis, affect the resulting material behavior. The partial dependence plot depicted in Fig. \ref{Figure:F4}B shows the effect of the mean nodal connectivity on the predicted effective stiffness. One can easily distinguish four regimes: two plateaus ($Z<3.7$ and $Z>5.1$) where change of nodal connectivity has a minimum effect on stiffness, separated by two regions ($3.7<Z<4.3$ and $4.3<Z<5.1$) where effective stiffness increases linearly, though with different slope, with regards to nodal connectivity. To fully comprehend these trends, it is constructive to compare them with Maxwell's criteria regarding the rigidity of pin-jointed frames \cite{Maxwell_Rule}. For the case of 2D structures, Maxwell's rule states that a nodal connectivity $Z=4$ is a necessary condition for rigidity, while the necessary and sufficient condition requires $Z=6$ \cite{Deshpande_2001_ActMat}. When applied to material microstructures, where strut junctions have resistance to rotation, these criteria can be used to classify cellular solids as stretching- or bending-dominated, referring to the governing internal load within their struts. The former class of materials is known to have a high stiffness that scales linearly with relative density (i.e. $\Bar{E}\sim \Bar{\rho}$) while bending-dominated lattices and foams are more flexible, with their stiffness following a cubic power-law (i.e. $\Bar{E}\sim \Bar{\rho}^3$). Our regression model can be used to rigorously derive these classifications and more importantly explain the behavior of those microstructures that do not fall within these categories. To do so, we first examine the power exponent $\beta$ of all cellular metamaterials in our dataset as a function of mean nodal connectivity (Fig. \ref{Figure:F4}C). It is clear that the four regimes identified by GAM are correlated, to a certain degree, with the trend shown here. That is, microstructures with very low nodal connectivity tend to be bending-dominated while high connectivity leads to stretching-dominated materials. In between the two regimes, stiffness increases significantly with nodal connectivity but at a decreasing rate, since it is gradually converging to the theoretical limit defined by the Hashin-Shtrikman bounds. However, there are two important deviations from Maxwell's rules: (i) one can notice that there are several materials with $Z<4$ that are nonetheless stretching-dominated, and (ii) it appears that $Z>4.5$ is sufficient to achieve rigidity in 2D metamaterials.

\begin{figure}[t]
\centering
\includegraphics[width=1.0\linewidth]{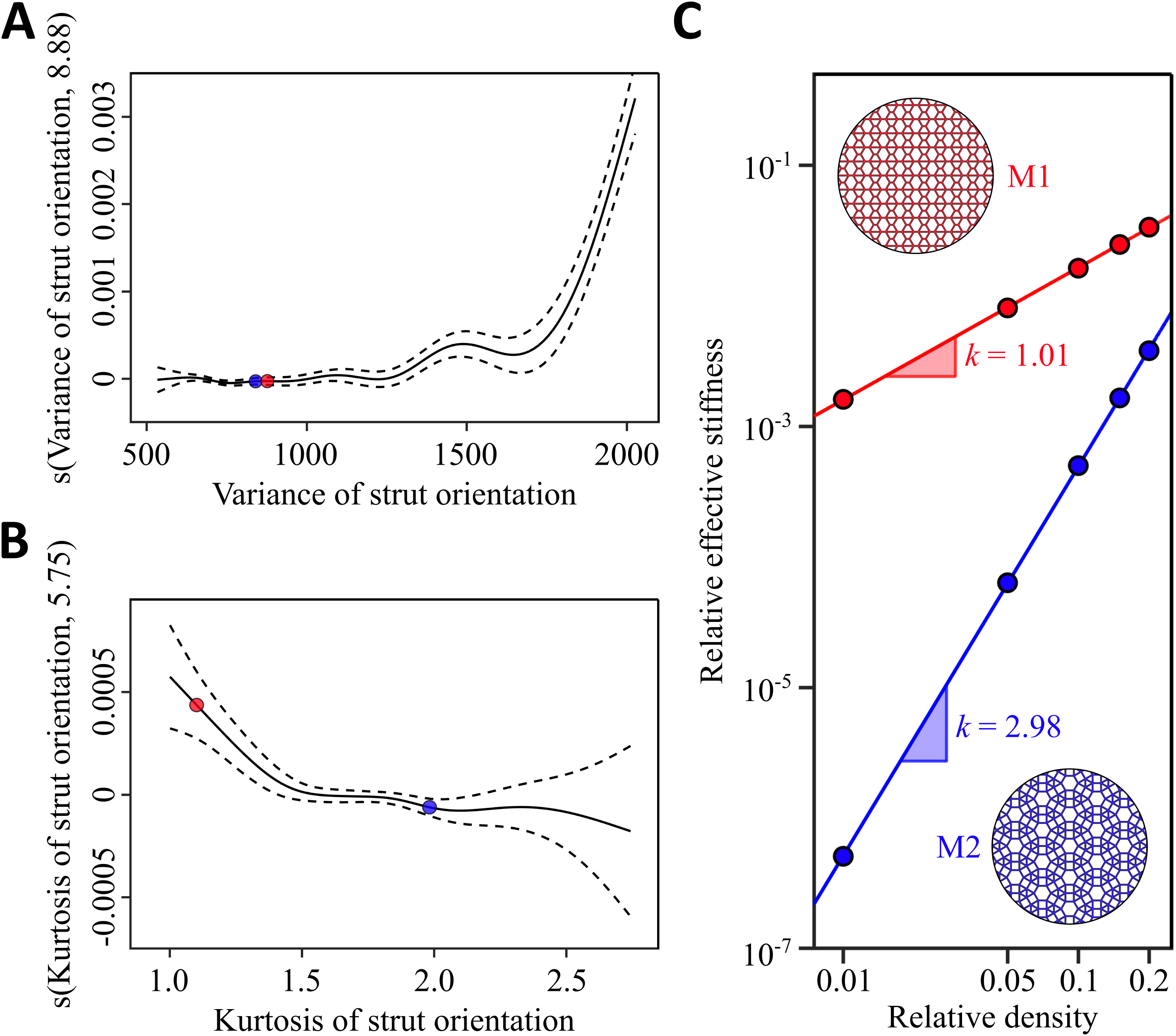}
\caption{Effect of strut orientation and behavior of two metamaterials that do not follow Maxwell's rules. (A) GAM partial dependence plot showing the relationship between effective stiffness and variance of strut orientation (M1 and M2 marked with red and blue dots). (B) GAM partial dependence plot showing the relationship between effective stiffness and kurtosis of strut orientation (M1 and M2 marked with red and blue dots). (C) The linear (M1) and cubic (M2) scaling of stiffness with density for the two cellular metamaterials.}
\label{Figure:F5}
\end{figure}

To understand why some structures violate Maxwell's rule, as applied to cellular metamaterials, attention needs to be given to the descriptors corresponding to strut orientation, as revealed by SHAP analysis. The corresponding GAM results for these descriptors are shown in Fig. \ref{Figure:F5}A-B. In both figures, one notices regions where the variance and/or the kurtosis of the strut orientation probability distribution, provide significant stiffness enhancement. To demonstrate this phenomenon, we choose two materials microstructures (corresponding to the outlier data shown in (Fig. \ref{Figure:F4}C)) whose mechanical behavior does not correlate to their mean nodal connectivity. The two metamaterials M1 and M2, shown in Fig. \ref{Figure:F5}C, have nodal connectivities $Z=3.5$ and $Z=4.5$ respectively. M1 is shown, however, to be much stiffer than M2. In addition, the effective stiffness of the material increases linearly with density, indicating its stretching-dominated behavior, even though it does not meet the criteria for rigidity according to Maxwell's rule i.e. $Z<4$. This discrepancy can be explained, in this case, through the large difference of the strut orientation kurtosis for M1 and M2, as highlighted in Fig. \ref{Figure:F5}B. Collectively, this analysis shows that nodal connectivity is insufficient to predict effective stiffness except for specific regimes, i.e. when $Z<3.3$ and $Z>4.5$. For all microstructures outside of that, one has to examine their struts' orientation, through the key descriptors, to understand the resulting material behavior. 

\section*{Discussion}

To conclude, we present here a data-driven framework for architected metamaterials that integrates virtual structure generation, microstructure quantification, machine-learning models and interpretability algorithms, in order to identify key morphological characteristics and their effects to effective stiffness. The results validate the importance of nodal connectivity on achieving the stiffest possible cellular metamaterials for a given relative density. It is further shown how strut orientation, represented through the second and fourth moments of its probability distribution, can become the critical factor that governs effective stiffness when the average nodal connectivity is not approaching its extremal values. Collectively, the findings demonstrate the ability of the developed framework to reveal structure-property relations that are inaccessible by conventional experimental and numerical techniques due to the vast number of involved parameters. It is important to highlight that even though effective stiffness is chosen here as a target property of interest, this data-driven approach can be applied to analyze any material property that can be calculated with reasonable accuracy, which for certain nonlinear problems may increase substantially the computational cost associated with the training and validation of the surrogate model. Furthermore, in cases where material properties sensitive to local flaws are investigated, additional morphological features and/or processing parameters (e.g. the resolution of the 3D-printer) should be considered as structure descriptors. It would be of immense interest to assess how the individual influence of each descriptor, represented here through the corresponding SHAP values, changes when different effective properties are examined. The proposed approach can therefore be further extended for the design of multifunctional metamaterials with tailored combinations of mechanical, thermal and/or acoustic properties. By modifying accordingly the structure vector, this framework can also be extended to other classes of metamaterials including architectures with curved members, thin shells, and/or density gradients. Finally, we envision that online databases that contain complete microstructure vectors and their corresponding material properties will greatly accelerate the design and discovery of novel architected metamaterials.

\matmethods{The compressive numerical simulation of the metamaterials is conducted using Abaqus (SIMULIA). Each strut is discretized into 10 shear deformable beam elements and the parent solid material is modeled as linear elastic. A vertical displacement is applied on top while all nodes at the bottom are fixed. The effective stiffness is measured as the slope of the stress-strain response (up to 2\% strain). The specimens are synthesized using a stereolithography-based printer (Form 3 by Formlabs) with 50$\mathrm{\mu}\mathrm{m}$ layer thickness using a Rigid-10K resin with Young's modulus $E=10000\mathrm{MPa}$. The post-processing protocol involves washing in isopropyl alcohol for 15 minutes to clear off the liquid resin, followed by UV-curing for 120 minutes in $70^{\circ}\mathrm{C}$. The compressive experiment is carried out using an MTS Criterion Series 40 testing stage quasi-statically, and both the force and displacement are measured by the MTS load cell.}

\showmatmethods{} % Display the Materials and Methods section

\acknow{This work was supported by Johns Hopkins University and the National Science Foundation (NSF) under Award Number 2129825.}

\showacknow{} % Display the acknowledgments section

% Bibliography
\bibliography{pnas-sample}

\end{document}

%% Comment out or remove this line before generating final copy for submission; this will also remove the warning re: "Consecutive odd pages found".

%% Adds the main heading for the SI text. Comment out this line if you do not have any supporting information text.
\SItext

\section*{1. Expanding basic designs to a large microstructural dataset}

\subsection*{A. Dual tessellations}
An obvious limitation of all \emph{k}-uniform tilings is that they consist solely of regular polygons, resulting in a lack of microstructural diversity that obviously affects any effort towards unbiased structure-property correlation. Hence, the concept of dual tessellation (or tiling) is applied to generate cellular architectures that, though still ordered, include numerous types of irregular polygons. Using power diagrams, any tiling is determined by the seed locations and their corresponding weights. To obtain the seed pattern of the dual of a tiling, a seed is placed at each of its vertices. This process is shown in Fig. S1 for three tilings (one regular, one semi-regular, and one 2-uniform) and their associated duals. It should be noted that creating the dual of a uniform weight-based tessellation leads to a regular tiling as well. In contrast, the dual tilings of multi-valued weight tessellations consist of irregular polygons. In the case of disordered cellular tessellations, their duals remain disordered but with polygons displaying sharper angles and lower compactness (see Fig. 1A).

\subsection*{B.Structure modification operations}
Despite having a primitive set of topologies that covers all well-studied 2D cellular materials, its microstructural diversity is still not sufficient to fully represent a broad class of cellular metamaterials. Hence, structure modification operations, including seed translation, increase of local weight variance, and modification of global weight variance, are required to enrich the microstructural information enclosed within the dataset. This process leads to the widening of the associated morphological descriptors' distributions that, as shown next, span continuously the space between their extremal values. Seed translation, as shown in Fig. S2A, leads to a modification of the polygonal tessellations by moving each seed to a new position, randomly placed within a predefined circle. Note that as the disorder of the initial topology increases, the effect of seed translation on the resulting morphological characteristics becomes minimal. The increase of local weight variance, as shown in Fig. S2B, modifies the distribution of individual seed weights to a wider normal distribution, thus reshaping and resizing each cell. The modification of global weight variance, as shown in Fig. S2C, also affects the seed weight, thereby resizing and reshaping the polygonal cells too, but in a macroscopically homogeneous manner. For example, if all seed weights in a semi-regular topology are switched to their mean value, the resulting microstructural becomes regular. It is readily obvious that modifying global weight variance has no impact on tessellations with single-valued seed weights. 
Since the aforementioned structure modification techniques are not equally applicable, or effective, to all microstructures, caution is needed to carefully tailor the coefficients involved in each operation to avoid inducing bias in the expanded dataset. Here, ordered tilings with single-valued seed weights (i.e., regular tilings and their duals) are subject to seed translation and an increase of local weight variance, each with five different magnitudes. The ordered tessellations with multi-valued seed weights (i.e., the semi-regular and 2-uniform tilings) are subject to all three operations, with three different magnitudes in each one. Disordered topologies (i.e., Voronoi-Laguerre) are generated twenty times with different local weight variance, and subsequently, their corresponding duals are created. The resulting expanded dataset consists of 1646 different topologies with a wide range of microstructural information.

\section*{2. Extracting morphological descriptors}

\subsection*{A. Morphological features}
Microstructure quantification for all 2D cellular metamaterials here is enabled by the use of physical morphological characteristics. Eleven features are selected and listed in Fig. S3 with their corresponding definitions. These features can be classified into four groups according to the relevant length-scale: (i) Macroscopic features, including relative density and number of cells, representing macro-structural information. (ii) At the cell level, we extract features that define the shape of each cell (or pore) including edge number, area, compactness, and eccentricity. (iii) Microscopic features correspond to characteristics of struts and junctions such as edge length, orientation, and nodal connectivity. (iv) At the intermediate level, we list features (distance and angle between neighboring seeds) that represent microstructural dispersity.

\subsection*{B. Feature distributions and descriptors}
Examining the selected features shows that only the macroscopic ones are deterministic, while all other features are characterized by a statistical distribution across the material domain. For the latter, we extract the first four moments (mean, variance, skewness, and kurtosis) of their distribution as representative descriptors of each feature. In contrast, the deterministic values of the macroscopic features correspond to two descriptors. This results in a structure vector $\mathbf{C}$ with 42 components in total. This descriptor vector $\mathbf{C}$ is calculated for all microstructures in our dataset, and the distribution of each descriptor among the dataset is shown in Fig. S4. These distributions exhibit a wide range, supporting both the representativity of our microstructure dataset as well as the effectiveness of feature-based microstructure quantification.

\section*{3. Machine learning modeling}

\subsection*{A. Feature selection techniques}
Pearson correlation and mutual information are employed to reduce the inherent multicollinearity of the microstructure descriptors. Here, the Pearson correlation coefficients between descriptors are first calculated as shown in Fig. S5. A threshold value $\lvert p \rvert=0.9$ is selected and two pairs of highly correlated features are detected. Hence, one descriptor from each pair (average cell edge number and average cell area) is removed, reducing the number of components in the structure vector to 40. Consequently, the mutual information metric is applied to examine the dependency between target property and  remaining microstructure descriptors. A threshold value $I=0$ is chosen and 17 descriptors uncorrelated with the target property are removed, including the number of polygon cells, all information regarding neighboring seed distance and cell area, the mean and kurtosis of strut length, the mean and skewness of neighboring seed angles, the skewness and kurtosis of cell eccentricity, and mean of strut orientation. Hence, the final reduced descriptor vector now has 23 components.

\subsection*{B. SHAP analysis for specific metamaterials}
By combining random forest modeling with SHAP analysis we have identified key microstructure features and their effect on effective stiffness. Here, we further utilize SHAP to interpret the prediction of individual microstructures of interest, to further elucidate structure-property relations. First, we focus on isotropic metamaterials that attain a stiffness close to the theoretical Hashin-Shtrikman upper bound, as shown in Fig. S6A. According to the SHAP result, this is mainly attributed to the extreme value of average nodal connectivity $Z=6$. Interestingly, the zero value of the variance of cell edge number is identified as an important contributor as well, implying that the stiffest microstructures will not exhibit any meaningful disorder. The commonly used square and diamond lattices, shown in Fig. S6B, are examined next. Based on the SHAP results, it can be seen that despite their equal mean nodal connectivity, the difference in their variance of strut orientation leads to a significant discrepancy in corresponding stiffness. This further emphasizes the importance of strut orientation in metamaterials with a moderate average nodal connectivity (i.e. within the two intermediate regimes identified by GAM). Lastly, we focus on a comparison between the triangular and Kagome lattices, which have been extensively studied, in part due to their shared theoretical stiffness-density relation (i.e., $\Bar{E}=\Bar{\rho}/3$). As shown in Fig. S6C, although the Kagome lattice has a much lower average nodal connectivity, with an associated smaller contribution to overall stiffness, the kurtosis of its strut orientation distribution provides an alternative positive effect that results in the Kagome lattice attaining a comparable rigidity to the triangular one. It should be noted that despite the equal stiffness predicted by beam theory, in practice the Kagome lattice will always display a lower stiffness, mainly due to the shear stresses and boundary effects.

\begin{figure}
\centering
\includegraphics[width=0.9\textwidth]{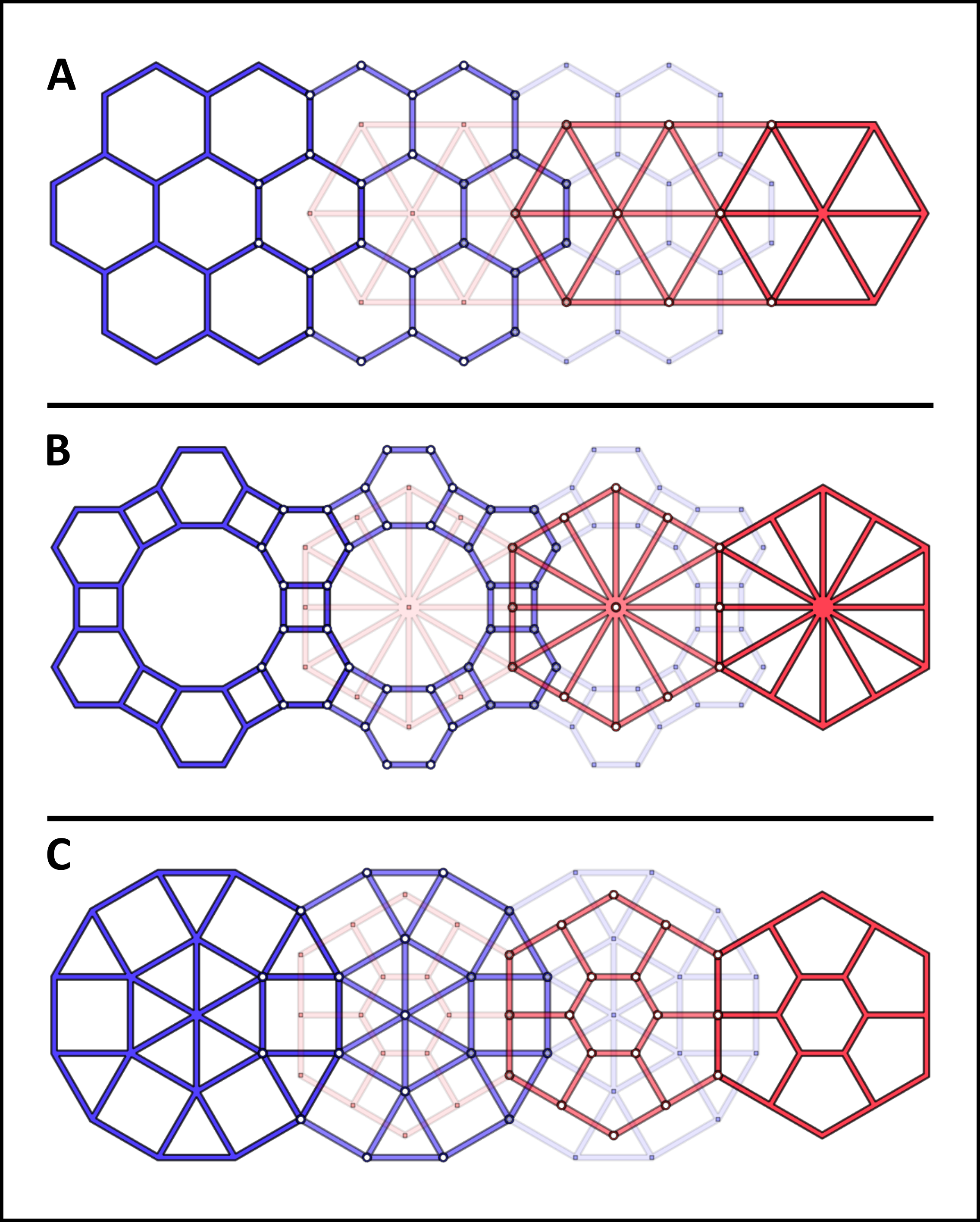}
\caption{Dual tessellations for (A) regular, (B) semi-regular, and (C) 2-uniform tilings: blue and red lines/markers correspond to edges/vertices of the original and dual tessellations respectively.}
\end{figure}

\begin{figure}
\centering
\includegraphics[width=1.0\textwidth]{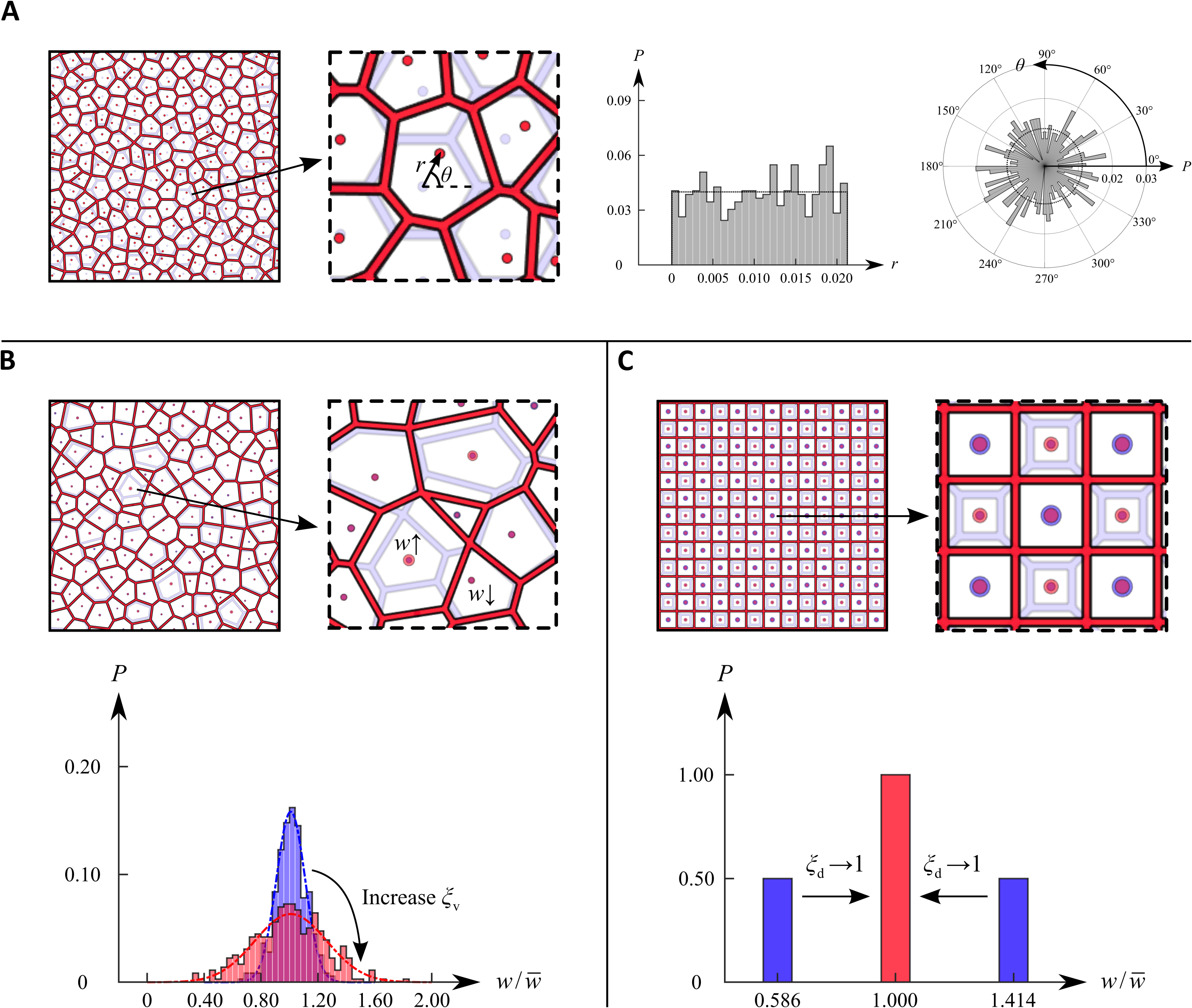}
\caption{Structure modification operations: (A) seed translation, (B) increase of local weight variance, and (C) modification of global weight variance. The blue color represents the components (seed, strut, distribution, etc.) in the original topologies, while the red color represents the corresponding components in the modified structures.}
\end{figure}

\begin{figure}
\centering
\includegraphics[width=0.8\textwidth]{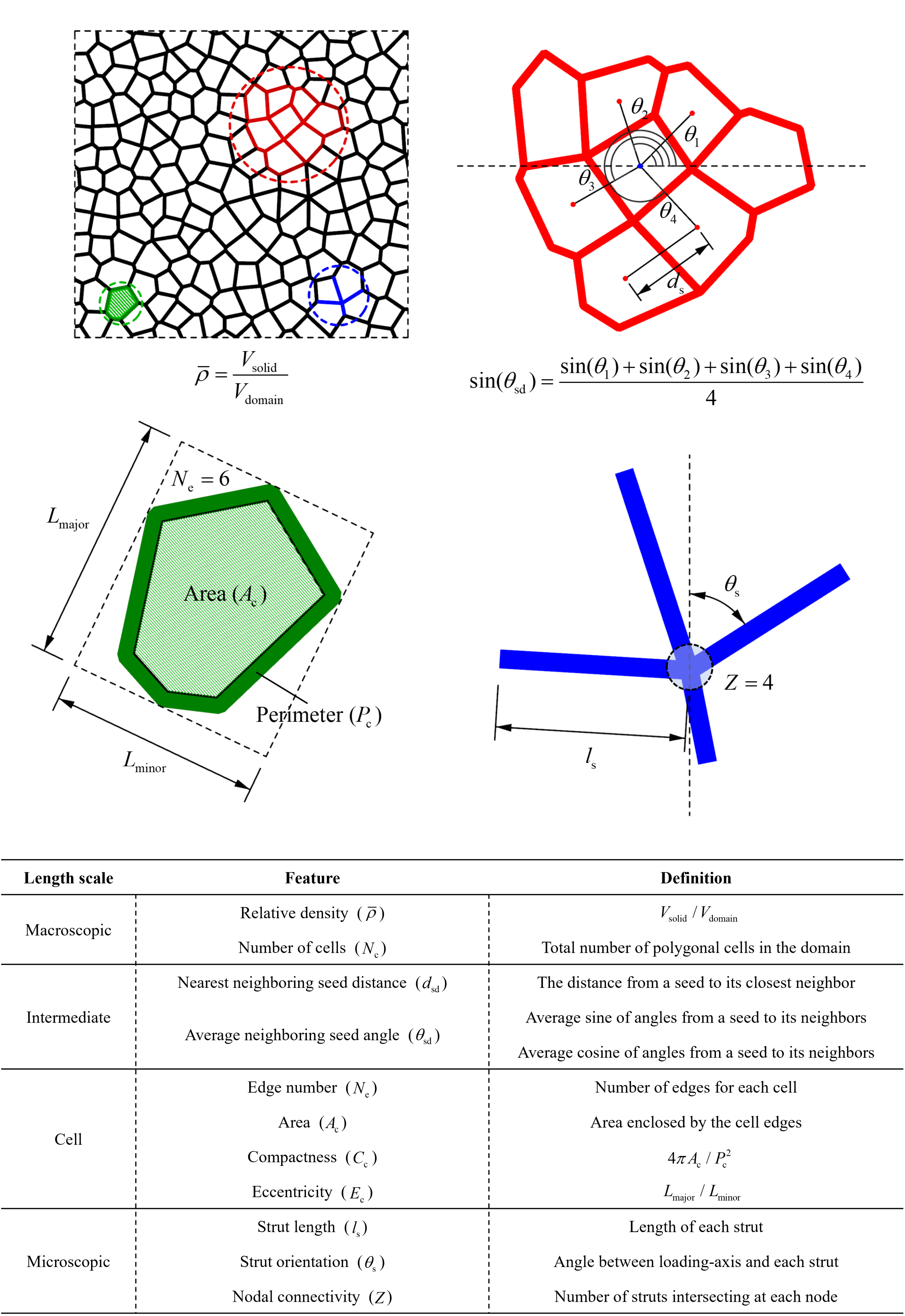}
\caption{Microstructural features.}
\end{figure}

\begin{figure}
\centering
\includegraphics[width=0.95\textwidth]{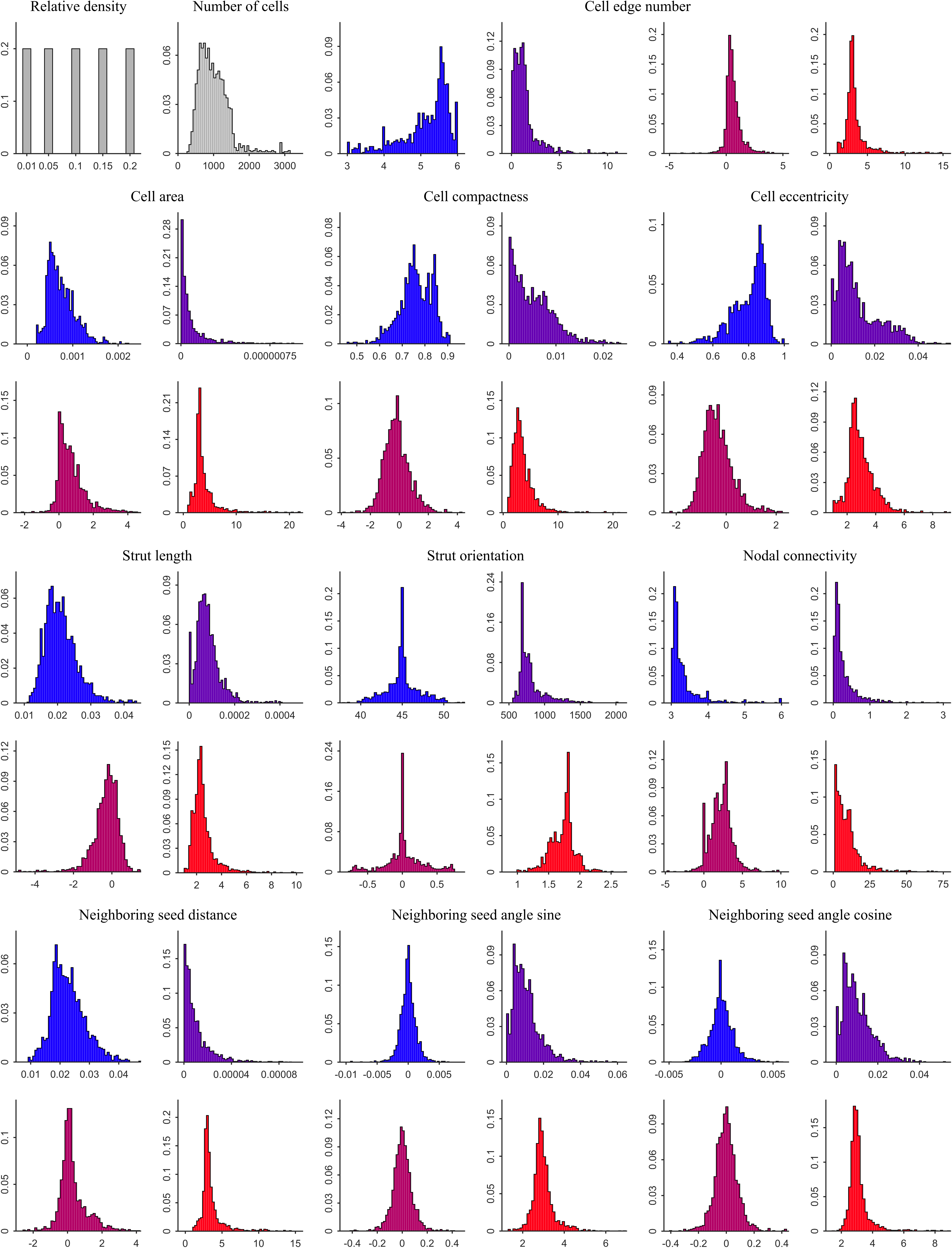}
\caption{Distribution of each morphological descriptor among all microstructures in our dataset. The deterministic features are represented by grey color, while the first to fourth moments of the statistical features are denoted by their color, changing from blue to red gradually. $x$-axis represents the feature value and $y$-axis denotes the probability.}
\end{figure}

\begin{figure}
\centering
\includegraphics[width=1.0\textwidth]{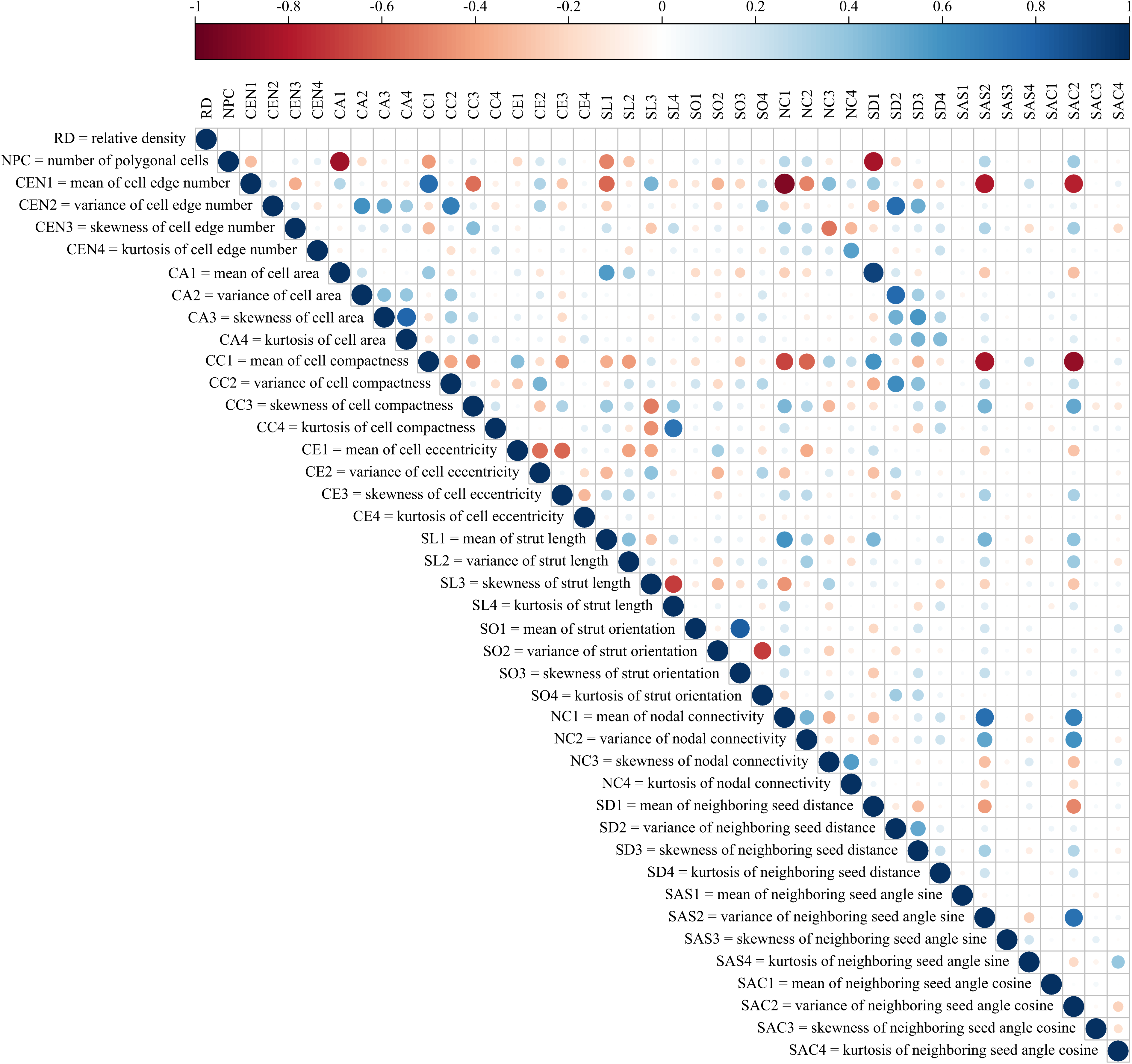}
\caption{The upper triangular map of Pearson correlation coefficient for all features in our microstructure dataset. The size of each circle represents the absolute value of the correlation coefficient.}
\end{figure}

\begin{figure}
\centering
\includegraphics[width=1.0\textwidth]{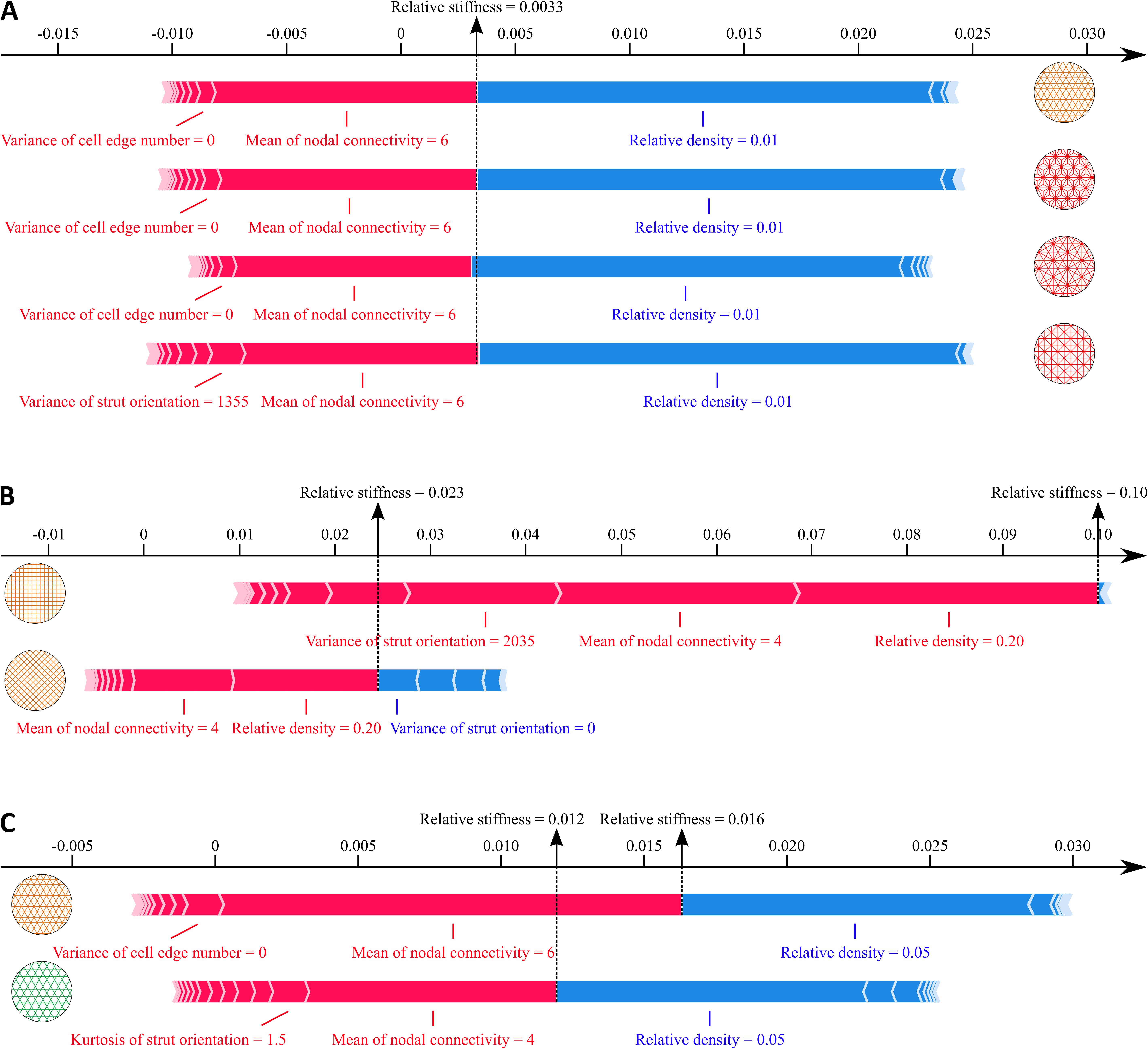}
\caption{SHAP force plots for three cases: (A) stiffest 2D metamaterials, (B) square and diamond lattices, and (C) triangular and Kagome lattices. Blue color represents negative contribution and red color denotes  positive contribution. The color in the metamaterial sketch refers to the classification in Fig. 1A.}
\end{figure}

%%% Add this line AFTER all your figures and tables
\FloatBarrier

\bibliography{pnas-sample}